# Analyzing negativity criterion in qubit-qutrit system located in external magnetic fields with the Dzyaloshinskii-Moriya interaction

Seyed Mohsen Moosavi Khansari[1]

*Department of Physics, Faculty of Basic Sciences, Ayatollah Boroujerdi University, Boroujerd, IRAN*

Fazlollah Kazemi Hasanvand[2]

*Department of Physics, Faculty of Basic Sciences, Ayatollah Boroujerdi University, Boroujerd, IRAN*

This investigation delves into the intricate entanglement dynamics exhibited by qubit-qutrit systems when subjected to the intricate dynamics of both the XXX isotropic Heisenberg model and the anisotropic XYZ model. These models are enriched by the presence of Dzyaloshinskii-Moriya (DM) interactions and external magnetic fields, creating a complex interplay of quantum phenomena. The exploration commences by initializing the system in a superposition of spin-coherent states, setting the stage for a nuanced analysis of entanglement evolution over time. By employing the negativity metric as the pivotal criterion for assessing entanglement, we gain insights into the subtle relationships between DM interactions, external magnetic fields, and the entanglement dynamics within this composite system. Through this comprehensive investigation, we aim to unravel the intricate dynamics and behaviors that emerge from the interplay of these fundamental elements, shedding light on the underlying principles governing quantum entanglement in this context.

## 1 Introduction

Entanglement stands as one of the most remarkable phenomena in quantum mechanics, playing a crucial role in information theory and quantum computing [1-5]. Coherent states are often regarded as quasi-classical due to their minimal uncertainty [6, 7], yet their combinations can generate strong quantum correlations, underscoring the non-classical nature of these states [7, 8]. Furthermore, the organic compound $ACu(PbaOH)(H_2O)_3.2H_2O$ exhibits two spins spin-1 and spin-1/2 in each unit cell [14, 16], making the study of entanglement in spin qubit-qutrit systems both relevant and necessary. Exploring the dynamic behavior of such systems can provide valuable insights into quantum correlations and their evolution over time. In this study, we investigated the entanglement dynamics of a qubit-qutrit system influenced by Dzyaloshinskii-Moriya (DM) interaction [17, 18], in the presence of an external magnetic field. Understanding how these interactions affect quantum states is vital for the progress of quantum information processing, quantum communication, and cryptography. The use of the negativity criterion as a measure of entanglement offers a powerful approach to quantifying quantum correlations, which serves as a resource for many quantum protocols. The qubit-qutrit system, consisting of a two-level and a three-level quantum entity, provides a rich platform for probing fundamental quantum phenomena. Introducing external magnetic fields and DM interaction adds complexity and realism to the model, as these factors can significantly modify the entanglement properties and quantum behavior of the system [9-13].

Significance of the Dzyaloshinskii-Moriya Interaction: This interaction is known for stabilizing spin textures and enabling topological phenomena, potentially supporting unconventional forms of entanglement and giving rise to novel quantum phases. Studying its influence on the negativity criterion can reveal important insights into the dynamics of entangled states under realistic conditions.
Practical Relevance: As quantum technologies advance, understanding how environmental factors like magnetic fields influence quantum states becomes crucial. Analyzing negativity in qubit-qutrit systems helps develop strategies to mitigate decoherence and preserve entanglement key steps toward robust quantum networks and devices.
Expanding Quantum Theoretical Foundations: Investigating the negativity criterion in coupled qubit-qutrit systems probes deeper into the nature of quantum correlations and non-locality. Such research can also inspire new mathematical frameworks for entanglement quantification, broadening our theoretical understanding and paving the way for innovative quantum applications.

[1] E-mail of corresponding author: m.moosavikhansari@abru.ac.ir

[2] E-mail: fa_kazemi270@yahoo.com

Through this research, we aim to advance both fundamental and practical knowledge in quantum science. The interplay between negativity, external magnetic fields, and DM interaction in qubit-qutrit systems holds promise for significant discoveries and technological developments in the quantum domain.

Outline of the Article: The first section is dedicated to essential theoretical calculations necessary for the subsequent analysis. The second section describes the initial state of the qubit-qutrit system and details the evolution governed by the Hamiltonian introduced earlier. Following this, an in-depth examination of the negativity related to the system's interactions is presented. The next segment discusses the results and their implications, culminating in a conclusion that summarizes key insights and future prospects.

## 2 Theoretical calculations

We examine a composite system of qubit and qutrit interacting with Dzyaloshinskii-Moriya (DM), each experiencing its own external magnetic field in both the isotropic $XXX$ and anisotropic $XYZ$ models. The Hamiltonian for this problem can be expressed as follows

$$H = \frac{1}{2}\sigma_{\mathrm{qb}}^{z} B_{z,\mathrm{qb}} + S_{qt}^{z} B_{z,\mathrm{qt}} + \frac{1}{2} D_{z}\left(\sigma_{\mathrm{qb}}^{x} S_{qt}^{y} - \sigma_{\mathrm{qb}}^{y} S_{qt}^{x}\right) + \frac{1}{2}\left(J_{x}\sigma_{\mathrm{qb}}^{x} S_{qt}^{x} + J_{y}\sigma_{\mathrm{qb}}^{y} S_{qt}^{y} + J_{z}\sigma_{\mathrm{qb}}^{z} S_{qt}^{z}\right) \quad (1)$$

in this regard, $B_{z,qb}$ is the external magnetic field applied to the qubit and $B_{z,qt}$ is the external magnetic field applied to the qutrit. $\sigma_{\mathrm{qb}}^{x}$ and $\sigma_{\mathrm{qb}}^{y}$ and $\sigma_{\mathrm{qb}}^{z}$ are the Pauli operators associated with the qubit. $S_{qt}^{x}$ and $S_{qt}^{y}$ and $S_{qt}^{z}$ are the operators related to the qutrit with spin $S = 1$. $D_z$ is the $z$ component of Dzyaloshinskii-Moriya coefficients. $J_i$ with $i = x, y, z$ show the strength of spin qubit-qutrit interaction. If $J_x = J_y = J_z$, we will have $XXX$ isotropic Heisenberg models, and if $J_x \neq J_y \neq J_z$, we will have $XYZ$ anisotropic Heisenberg model. In quantum information theory, negativity is a key measure of entanglement that provides insights into the properties of quantum states. Below is a detailed summary of the methods of negativity calculation based on the synthesized information sourced from various studies and articles. To calculate quantum entanglement, we use negativity scale. Negativity for a quantum state with density matrix $\rho$ is defined as [19-24]:

$$N(\rho) = \frac{1}{2}(||\rho^{T_i}|| - 1) \quad (2)$$

Here $||\rho^{T_i}||$ refers to the trace norm, i.e., the sum of singular values of $\rho^{T_i}$, $\rho^{T_i}$ represents the partial transpose of $\rho$ with respect to the specific component labeled as $i$. Negativity is an important and increasingly utilized measure of quantum entanglement. It provides insights into the correlations within quantum systems and aids in the analysis of quantum states through their density matrices. Negativity is not only a theoretical construct but has practical implications in quantum metrology, quantum computing, and communication. For example, quantum negativity has been shown to increase measurement precision in quantum systems, enhancing technologies such as quantum computers and photonic quantum technologies. In some cases, negativity is computed using geometric methods by analyzing the convex structure of the set of separable states. The negativity can then be viewed geometrically as a quantity related to the distances between quantum states and the set of separable states. Various numerical approaches can also be employed for calculating negativity, especially for higher-dimensional systems or mixed states where analytical solutions become cumbersome. Techniques like the method of moments or the semi-definite programming can be used to effectively compute negativity [22,25,26].

## 3 The Spin coherent states serve as the initial state for the qubit-qutrit system

The spin coherence state is introduced as follows:

$$|\alpha, j\rangle = (|\alpha|^2 + 1)^{-j} \sum_{m=-j}^{j} \sqrt{\binom{2j}{j+m}} \alpha^{j+m} |j, m\rangle \tag{3}$$

the coherent state of the qubit is obtained by setting $j = 1/2$ as follows

$$|\alpha, \tfrac{1}{2}\rangle = \frac{\left|\frac{1}{2}, -\frac{1}{2}\right\rangle}{\sqrt{|\alpha|^2+1}} + \frac{\alpha\left|\frac{1}{2}, \frac{1}{2}\right\rangle}{\sqrt{|\alpha|^2+1}} \tag{4}$$

with the definitions $\left|\frac{1}{2}, -\frac{1}{2}\right\rangle \to |0\rangle$ and $\left|\frac{1}{2}, \frac{1}{2}\right\rangle \to |1\rangle$, we can write

$$|\alpha, \tfrac{1}{2}\rangle = \frac{|0\rangle}{\sqrt{|\alpha|^2+1}} + \frac{\alpha|1\rangle}{\sqrt{|\alpha|^2+1}} \tag{5}$$

The coherent state of the qutrit is obtained by setting $j = 1$ as follows

$$|\alpha, 1\rangle = \frac{\alpha^2|1,1\rangle}{|\alpha|^2+1} + \frac{\sqrt{2}\alpha|1,0\rangle}{|\alpha|^2+1} + \frac{|1,-1\rangle}{|\alpha|^2+1} \tag{6}$$

with the definitions $|1,-1\rangle \to |0\rangle$ and $|1,0\rangle \to |1\rangle$ and $|1,1\rangle \to |2\rangle$, we can write

$$|\alpha, 1\rangle = \frac{\alpha^2|2\rangle}{|\alpha|^2+1} + \frac{\sqrt{2}\alpha|1\rangle}{|\alpha|^2+1} + \frac{|0\rangle}{|\alpha|^2+1} \tag{7}$$

We create a pure entanglement state by superposing spin coherent states of the qubit and qutrit

$$\begin{aligned}
&|\psi(0)\rangle = \\
&|00\rangle\left(\frac{\cos(\theta)}{(|\alpha_1|^2+1)\sqrt{|\beta_1|^2+1}} + \frac{e^{-i\varphi}\sin(\theta)}{(|\alpha_2|^2+1)\sqrt{|\beta_2|^2+1}}\right) + \\
&|01\rangle\left(\frac{\sqrt{2}\alpha_1\cos(\theta)}{(|\alpha_1|^2+1)\sqrt{|\beta_1|^2+1}} + \frac{\sqrt{2}\alpha_2 e^{-i\varphi}\sin(\theta)}{(|\alpha_2|^2+1)\sqrt{|\beta_2|^2+1}}\right) + \\
&|02\rangle\left(\frac{\alpha_1^2\cos(\theta)}{(|\alpha_1|^2+1)\sqrt{|\beta_1|^2+1}} + \frac{\alpha_2^2 e^{-i\varphi}\sin(\theta)}{(|\alpha_2|^2+1)\sqrt{|\beta_2|^2+1}}\right) + \\
&|10\rangle\left(\frac{\beta_1\cos(\theta)}{(|\alpha_1|^2+1)\sqrt{|\beta_1|^2+1}} + \frac{\beta_2 e^{-i\varphi}\sin(\theta)}{(|\alpha_2|^2+1)\sqrt{|\beta_2|^2+1}}\right) +
\end{aligned}$$

$$|11\rangle\left(\frac{\sqrt{2}\alpha_1\beta_1\cos(\theta)}{(|\alpha_1|^2+1)\sqrt{|\beta_1|^2+1}}+\frac{\sqrt{2}\alpha_2\beta_2e^{-i\varphi}\sin(\theta)}{(|\alpha_2|^2+1)\sqrt{|\beta_2|^2+1}}\right)+$$
$$|12\rangle\left(\frac{\alpha_1^2\beta_1\cos(\theta)}{(|\alpha_1|^2+1)\sqrt{|\beta_1|^2+1}}+\frac{\alpha_2^2\beta_2e^{-i\varphi}\sin(\theta)}{(|\alpha_2|^2+1)\sqrt{|\beta_2|^2+1}}\right) \tag{8}$$

We consider $\theta=\frac{\pi}{4},\varphi=0$ and $\alpha_1=\beta_1=\alpha$ and $\alpha_2=\beta_2=-\alpha$, the discussed state can now be written as follows

$$|\psi(0)\rangle=\frac{\sqrt{2}\alpha^2|02\rangle}{(|\alpha|^2+1)^{3/2}}+\frac{2\alpha^2|11\rangle}{(|\alpha|^2+1)^{3/2}}+\frac{\sqrt{2}|00\rangle}{(|\alpha|^2+1)^{3/2}} \tag{9}$$

Using this state, it is possible to calculate the normalization and the density operator relation of the initial state of the system in Dirac notation as follows

$$\langle\psi(0)|\psi(0)\rangle=\frac{6\alpha^4}{(|\alpha|^2+1)^3}+\frac{2}{(|\alpha|^2+1)^3} \tag{10}$$

$$\rho(0)=\frac{2\alpha^4|02\rangle\langle02|}{(\alpha^2+1)^3}+\frac{2\sqrt{2}\alpha^4|11\rangle\langle02|}{(\alpha^2+1)^3}+\frac{2\sqrt{2}\alpha^4|02\rangle\langle11|}{(\alpha^2+1)^3}+\frac{4\alpha^4|11\rangle\langle11|}{(\alpha^2+1)^3}+\frac{2\alpha^2|02\rangle\langle00|}{(\alpha^2+1)^3}+\frac{2\sqrt{2}\alpha^2|11\rangle\langle00|}{(\alpha^2+1)^3}+$$
$$\frac{2\alpha^2|00\rangle\langle02|}{(\alpha^2+1)^3}+\frac{2\sqrt{2}\alpha^2|00\rangle\langle11|}{(\alpha^2+1)^3}+\frac{2|00\rangle\langle00|}{(\alpha^2+1)^3} \tag{11}$$

The Hamiltonian matrix arrays in the bases of this qubit and qutrit can be determined as follows

$$\begin{aligned}&H_{1,2}=H_{1,3}=H_{1,4}=H_{1,6}=H_{2,1}=\\&H_{2,3}=H_{2,5}=H_{3,1}=H_{3,2}=H_{3,4}=\\&H_{3,6}=H_{4,1}=H_{4,3}=H_{4,5}=H_{4,6}=\\&H_{5,2}=H_{5,4}=H_{5,6}=H_{6,1}=H_{6,3}=\\&H_{6,4}=H_{6,5}=0\end{aligned} \tag{12}$$

$$H_{1,1}=\frac{1}{2}\left(B_{z,\mathrm{qb}}+2B_{z,\mathrm{qt}}+J_z\right),\quad H_{1,5}=\frac{J_x-J_y}{2\sqrt{2}} \tag{13}$$

$$H_{2,2}=\frac{B_{z,\mathrm{qb}}}{2},\quad H_{2,4}=\frac{2iD_z+J_x+J_y}{2\sqrt{2}},\quad H_{2,6}=\frac{J_x-J_y}{2\sqrt{2}} \tag{14}$$

$$H_{3,3}=\frac{1}{2}\left(B_{z,\mathrm{qb}}-2B_{z,\mathrm{qt}}-J_z\right),\quad H_{3,5}=\frac{2iD_z+J_x+J_y}{2\sqrt{2}} \tag{15}$$

$$H_{4,2}=\frac{-2iD_z+J_x+J_y}{2\sqrt{2}},\quad H_{4,4}=-\frac{B_{z,\mathrm{qb}}}{2}+B_{z,\mathrm{qt}}-\frac{J_z}{2} \tag{16}$$

$$H_{5,1}=\frac{J_x-J_y}{2\sqrt{2}},\quad H_{5,3}=\frac{-2iD_z+J_x+J_y}{2\sqrt{2}},\quad H_{5,5}=-\frac{B_{z,\mathrm{qb}}}{2} \tag{17}$$

$$H_{6,2}=\frac{J_x-J_y}{2\sqrt{2}},\quad H_{6,6}=\frac{1}{2}\left(-B_{z,\mathrm{qb}}-2B_{z,\mathrm{qt}}+J_z\right) \tag{18}$$

With the Hamiltonian given, the state $|\psi(0)\rangle$ evolves under $U(t)=\exp(-iHt)$ as $|\psi(t)\rangle=U(t)|\psi(0)\rangle$. The time-dependent system's density operator is then $\rho(t)=|\psi(t)\rangle\langle\psi(t)|$.

## 4 Results and discussion

In the upcoming section, we aim to provide a comprehensive overview of the results we have garnered, delving into a detailed analysis and offering nuanced interpretations to enrich our understanding of the outcomes. Numerical methodologies and computational approaches have been consciously incorporated and utilized to accurately compute and analyze every aspect of the results in the study.

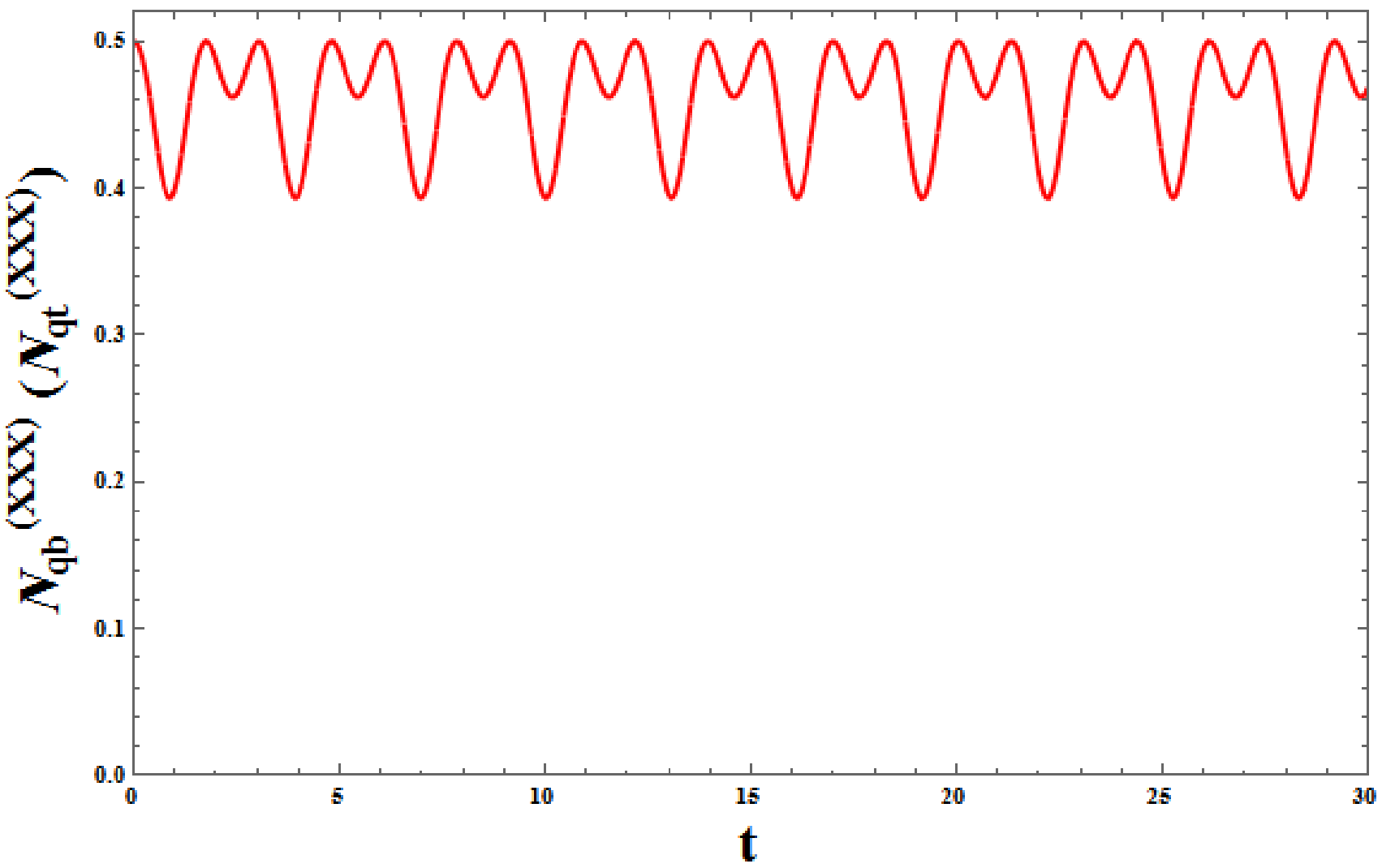

Figure 1: Time negativity diagram for the isotropic Heisenberg $XXX$ model for qubit (qutrit) subsystem is illustrated with initial values of $\alpha = 1$, $J_x = 1, J_y = 1, J_z = 1, B_{z,qb} = 1, B_{z,qt} = 1, D_x = 0, D_y = 0, D_z = 1, \varphi = 0, \theta = \pi/4$.

In Fig. 1 the diagram of negativity in terms of time is depicted in the Heisenberg isotropic model for the qubit (qutrit) subsystem. The negativity, or a measure equivalent to entanglement, exhibits dynamic variations as time progresses, showcasing the evolving nature of entangled states and their intricate properties over different temporal intervals. The negativity values span from 0.392 to 0.5, revealing a fascinating symmetry in the presence of dual peaks at the maximum negativity level. This distinctive pattern showcases the intriguing behavior of the system, mirroring itself at its peak negativity points, offering a captivating insight into the dynamics of the entangled states under examination. The computed average negativity value, derived through a meticulous analysis of the temporal progression, stands at a precise numerical value of 0.462, encapsulating the essence of the entanglement dynamics observed over the duration under scrutiny. The calculations were performed using the following relations

$$|\psi(0)\rangle = \frac{1}{2}|00\rangle + \frac{1}{2}|02\rangle + \frac{|11\rangle}{\sqrt{2}} \tag{19}$$

$$\rho(0) = \frac{1}{4}|00\rangle\langle 00| + \frac{1}{4}|02\rangle\langle 00| + \frac{|11\rangle\langle 00|}{2\sqrt{2}} + \frac{1}{4}|00\rangle\langle 02| + \frac{1}{4}|02\rangle\langle 02| + \frac{|11\rangle\langle 02|}{2\sqrt{2}} + \frac{|00\rangle\langle 11|}{2\sqrt{2}} + \frac{|02\rangle\langle 11|}{2\sqrt{2}} + \frac{1}{2}|11\rangle\langle 11| \tag{20}$$

$$H = \begin{pmatrix} 2 & 0 & 0 & 0 & 0 & 0 \\ 0 & \frac{1}{2} & 0 & \frac{1+2i}{\sqrt{2}} & 0 & 0 \\ 0 & 0 & -1 & 0 & \frac{1+2i}{\sqrt{2}} & 0 \\ 0 & \frac{1-2i}{\sqrt{2}} & 0 & 0 & 0 & 0 \\ 0 & 0 & \frac{1-2i}{\sqrt{2}} & 0 & -\frac{1}{2} & 0 \\ 0 & 0 & 0 & 0 & 0 & -1 \end{pmatrix} \tag{21}$$

$$\begin{aligned} U = {} & e^{-2it}|00\rangle\langle 00| + e^{it}|12\rangle\langle 12| + \\ & \left(\frac{(\sqrt{41}-1)e^{\frac{1}{4}i(\sqrt{41}-1)t}}{2\sqrt{41}} - \frac{(-1-\sqrt{41})e^{-\frac{1}{4}i(1+\sqrt{41})t}}{2\sqrt{41}}\right)|01\rangle\langle 01| + \\ & \left((-1+2i)\sqrt{\frac{2}{41}}e^{\frac{1}{4}i(\sqrt{41}-1)t} + (1-2i)\sqrt{\frac{2}{41}}e^{-\frac{1}{4}i(1+\sqrt{41})t}\right)|10\rangle\langle 01| + \\ & \left(\frac{(1+\sqrt{41})e^{\frac{1}{4}i(3+\sqrt{41})t}}{2\sqrt{41}} - \frac{(1-\sqrt{41})e^{-\frac{1}{4}i(\sqrt{41}-3)t}}{2\sqrt{41}}\right)|02\rangle\langle 02| + \\ & \left((1-2i)\sqrt{\frac{2}{41}}e^{-\frac{1}{4}i(\sqrt{41}-3)t} - (1-2i)\sqrt{\frac{2}{41}}e^{\frac{1}{4}i(3+\sqrt{41})t}\right)|11\rangle\langle 02| + \\ & \left((-1-2i)\sqrt{\frac{2}{41}}e^{\frac{1}{4}i(\sqrt{41}-1)t} + (1+2i)\sqrt{\frac{2}{41}}e^{-\frac{1}{4}i(1+\sqrt{41})t}\right)|01\rangle\langle 10| + \\ & \left(\frac{(1+\sqrt{41})e^{\frac{1}{4}i(\sqrt{41}-1)t}}{2\sqrt{41}} - \frac{(1-\sqrt{41})e^{-\frac{1}{4}i(1+\sqrt{41})t}}{2\sqrt{41}}\right)|10\rangle\langle 10| + \\ & \left((1+2i)\sqrt{\frac{2}{41}}e^{-\frac{1}{4}i(\sqrt{41}-3)t} - (1+2i)\sqrt{\frac{2}{41}}e^{\frac{1}{4}i(3+\sqrt{41})t}\right)|02\rangle\langle 11| + \\ & \left(\frac{(\sqrt{41}-1)e^{\frac{1}{4}i(3+\sqrt{41})t}}{2\sqrt{41}} - \frac{(-1-\sqrt{41})e^{-\frac{1}{4}i(\sqrt{41}-3)t}}{2\sqrt{41}}\right)|11\rangle\langle 11| \end{aligned} \tag{22}$$

$$\begin{aligned} |\psi(t)\rangle = {} & \frac{1}{2}e^{-2it}|00\rangle + \left(\frac{1}{4}e^{-\frac{1}{4}i(\sqrt{41}-3)t} + \frac{\left(\frac{3}{4}+2i\right)e^{-\frac{1}{4}i(\sqrt{41}-3)t}}{\sqrt{41}} + \frac{1}{4}e^{\frac{1}{4}i(3+\sqrt{41})t} - \frac{\left(\frac{3}{4}+2i\right)e^{\frac{1}{4}i(3+\sqrt{41})t}}{\sqrt{41}}\right)|02\rangle + \\ & \left(\frac{e^{-\frac{1}{4}i(\sqrt{41}-3)t}}{2\sqrt{2}} + \frac{\left(\frac{3}{2}-2i\right)e^{-\frac{1}{4}i(\sqrt{41}-3)t}}{\sqrt{82}} + \frac{e^{\frac{1}{4}i(3+\sqrt{41})t}}{2\sqrt{2}} - \frac{\left(\frac{3}{2}-2i\right)e^{\frac{1}{4}i(3+\sqrt{41})t}}{\sqrt{82}}\right)|11\rangle \end{aligned} \tag{23}$$

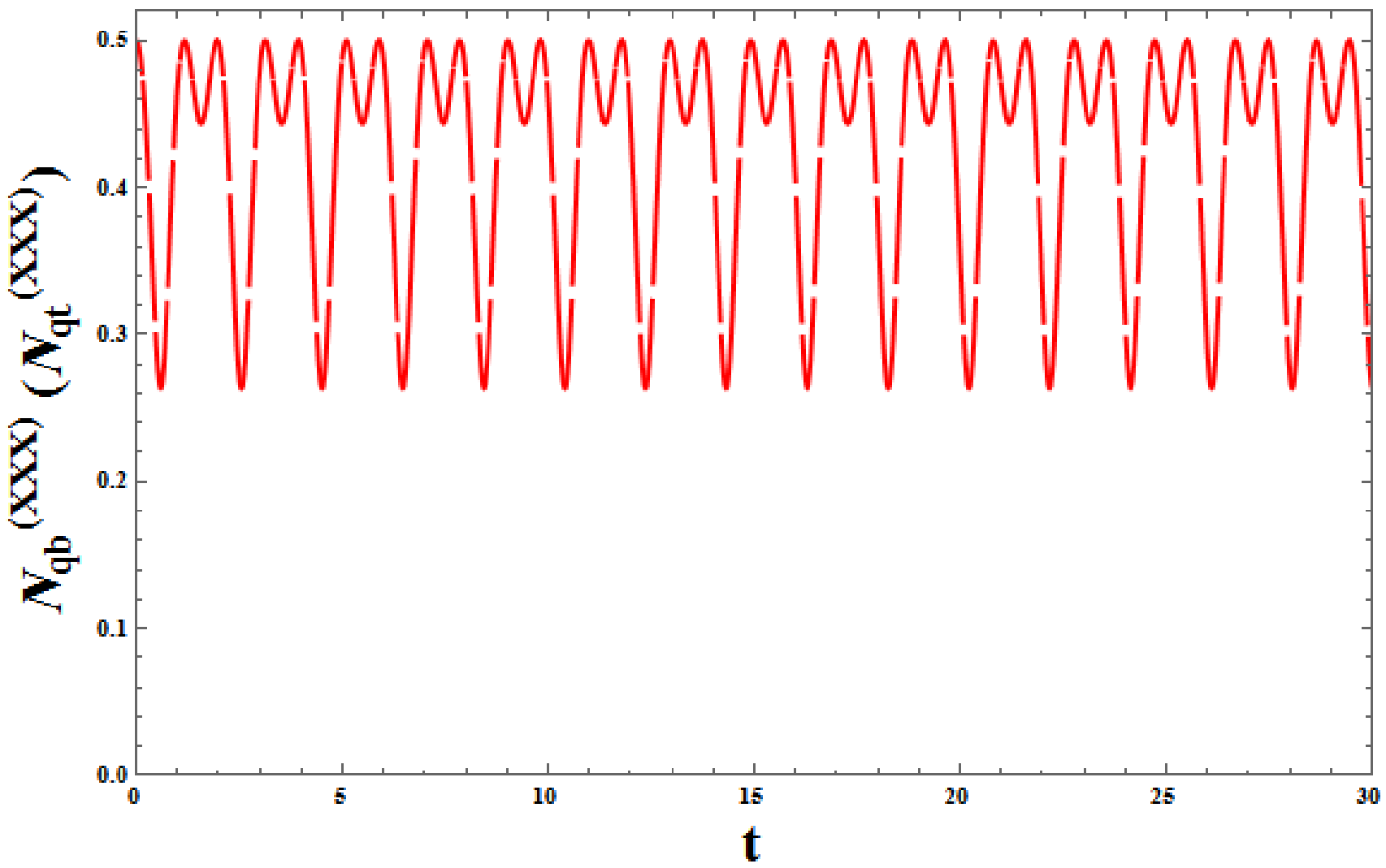

Figure 2: Time negativity diagram for the isotropic Heisenberg $XXX$ model for qubit (qutrit) subsystem is illustrated with initial values of $\alpha = 1$, $J_x = 1, J_y = 1, J_z = 1, B_{z,qb} = 1, B_{z,qt} = 1, D_x = 0, D_y = 0, D_z = 2, \varphi = 0, \theta = \pi/4$.

In Heisenberg's isotropic model, Fig. 2 depicts the negativity diagram over time with $D_z = 2$ for the qubit subsystem. Once more, it is worth noting that negativity, which is synonymous with entanglement, exhibits fluctuations throughout the progression of time. This variability in negativity underscores the dynamic nature of entanglement phenomena as they evolve over time intervals. Once again, it is observed that the maximum negativity reaches a value of 0.5, indicating the pinnacle of entanglement exhibited in this scenario. On the other end of the spectrum, the minimum negativity value in this case stands at 0.261, marking the lower threshold of entanglement manifestation within this specific context. These distinct numerical values serve as important benchmarks, delineating the range within which entanglement dynamics unfold and fluctuate. The time average of negativity is 0.425, indicating a decrease as the DM interaction strength increases.

The calculations were performed using the following relations

$$H = \begin{pmatrix} 2 & 0 & 0 & 0 & 0 & 0 \\ 0 & \frac{1}{2} & 0 & \frac{1+2i}{\sqrt{2}} & 0 & 0 \\ 0 & 0 & -1 & 0 & \frac{1+2i}{\sqrt{2}} & 0 \\ 0 & \frac{1-2i}{\sqrt{2}} & 0 & 0 & 0 & 0 \\ 0 & 0 & \frac{1-2i}{\sqrt{2}} & 0 & -\frac{1}{2} & 0 \\ 0 & 0 & 0 & 0 & 0 & -1 \end{pmatrix} \tag{24}$$

$$U = e^{-2it}|00\rangle\langle 00| + e^{it}|12\rangle\langle 12| +$$
$$\left(\frac{(\sqrt{41}-1)e^{\frac{1}{4}i(\sqrt{41}-1)t}}{2\sqrt{41}} - \frac{(-1-\sqrt{41})e^{-\frac{1}{4}i(1+\sqrt{41})t}}{2\sqrt{41}}\right)|01\rangle\langle 01| +$$
$$\left((-1+2i)\sqrt{\frac{2}{41}}e^{\frac{1}{4}i(\sqrt{41}-1)t} + (1-2i)\sqrt{\frac{2}{41}}e^{-\frac{1}{4}i(1+\sqrt{41})t}\right)|10\rangle\langle 01| +$$
$$\left(\frac{(1+\sqrt{41})e^{\frac{1}{4}i(3+\sqrt{41})t}}{2\sqrt{41}} - \frac{(1-\sqrt{41})e^{-\frac{1}{4}i(\sqrt{41}-3)t}}{2\sqrt{41}}\right)|02\rangle\langle 02| +$$

$$\left((1-2i)\sqrt{\frac{2}{41}}e^{-\frac{1}{4}i(\sqrt{41}-3)t}-(1-2i)\sqrt{\frac{2}{41}}e^{\frac{1}{4}i(3+\sqrt{41})t}\right)|11\rangle\langle 02|+$$
$$\left((-1-2i)\sqrt{\frac{2}{41}}e^{\frac{1}{4}i(\sqrt{41}-1)t}+(1+2i)\sqrt{\frac{2}{41}}e^{-\frac{1}{4}i(1+\sqrt{41})t}\right)|01\rangle\langle 10|+$$
$$\left(\frac{(1+\sqrt{41})e^{\frac{1}{4}i(\sqrt{41}-1)t}}{2\sqrt{41}}-\frac{(1-\sqrt{41})e^{-\frac{1}{4}i(1+\sqrt{41})t}}{2\sqrt{41}}\right)|10\rangle\langle 10|+$$
$$\left((1+2i)\sqrt{\frac{2}{41}}e^{-\frac{1}{4}i(\sqrt{41}-3)t}-(1+2i)\sqrt{\frac{2}{41}}e^{\frac{1}{4}i(3+\sqrt{41})t}\right)|02\rangle\langle 11|+$$
$$\left(\frac{(\sqrt{41}-1)e^{\frac{1}{4}i(3+\sqrt{41})t}}{2\sqrt{41}}-\frac{(-1-\sqrt{41})e^{-\frac{1}{4}i(\sqrt{41}-3)t}}{2\sqrt{41}}\right)|11\rangle\langle 11| \qquad (25)$$

$$|\psi(t)\rangle=\frac{1}{2}e^{-2it}|00\rangle+\left(\frac{1}{4}e^{-\frac{1}{4}i(\sqrt{41}-3)t}+\frac{\left(\frac{3}{4}+2i\right)e^{-\frac{1}{4}i(\sqrt{41}-3)t}}{\sqrt{41}}+\frac{1}{4}e^{\frac{1}{4}i(3+\sqrt{41})t}-\frac{\left(\frac{3}{4}+2i\right)e^{\frac{1}{4}i(3+\sqrt{41})t}}{\sqrt{41}}\right)|02\rangle+$$
$$\left(\frac{e^{-\frac{1}{4}i(\sqrt{41}-3)t}}{2\sqrt{2}}+\frac{\left(\frac{3}{2}-2i\right)e^{-\frac{1}{4}i(\sqrt{41}-3)t}}{\sqrt{82}}+\frac{e^{\frac{1}{4}i(3+\sqrt{41})t}}{2\sqrt{2}}-\frac{\left(\frac{3}{2}-2i\right)e^{\frac{1}{4}i(3+\sqrt{41})t}}{\sqrt{82}}\right)|11\rangle \qquad (26)$$

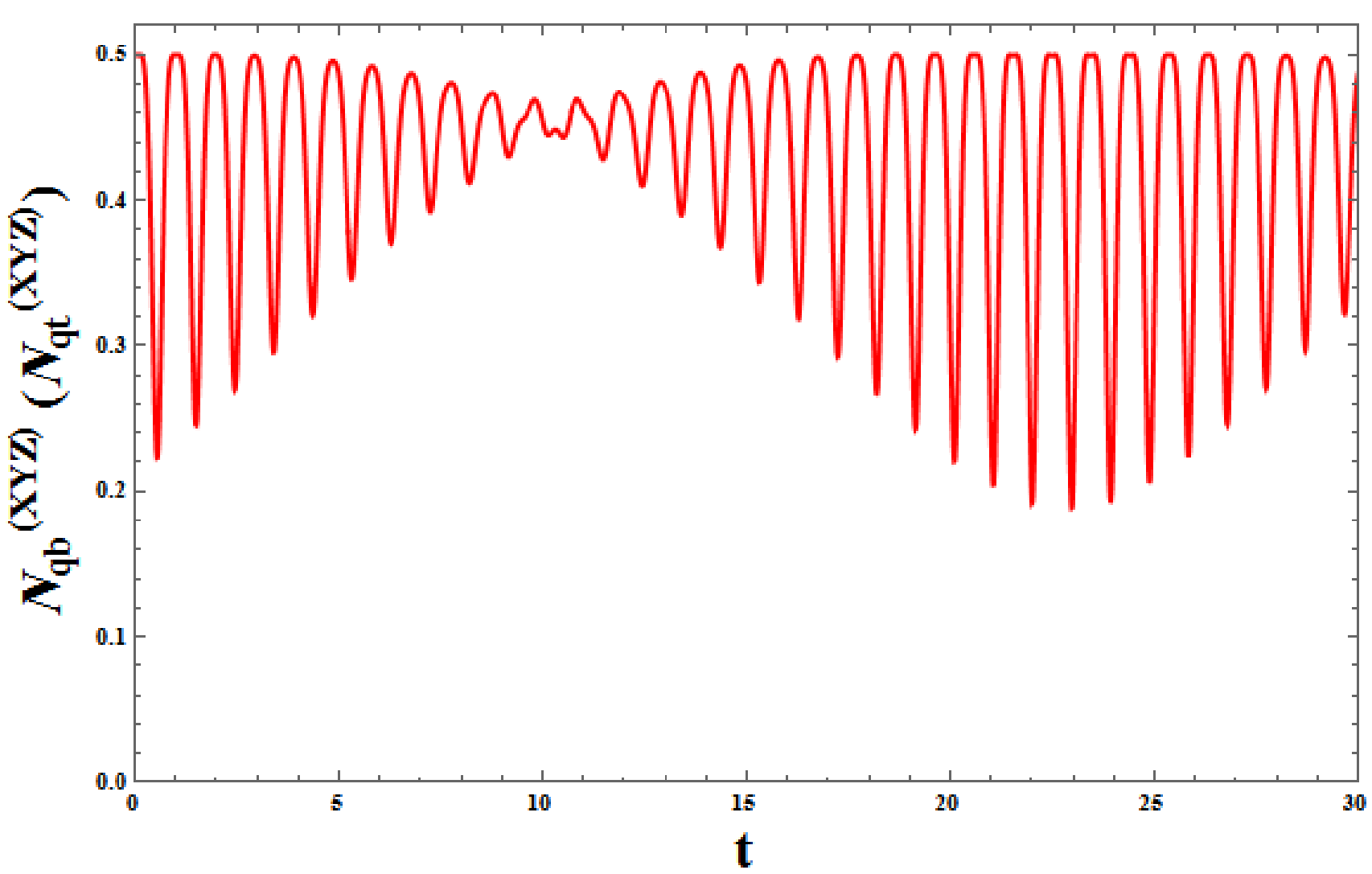

Figure 3: Time negativity diagram for the anisotropic Heisenberg $XYZ$ model for qubit (qutrit) subsystem is illustrated with initial values of $\alpha=1$, $J_x=1, J_y=5, J_z=10, B_{z,qb}=1, B_{z,qt}=1, D_x=0, D_y=0, D_z=1, \varphi=0, \theta=\pi/4$.

In Fig. 3, the negativity diagram in terms of time for the anisotropic Heisenberg $XYZ$ model for the qubit (qutrit) subsystem and the initial values of $J_x=1, J_y=5, J_z=10$ is depicted. The value representing the highest level of negativity in this scenario is 0.5, indicating a significant degree of adverse sentiment. Conversely, at the other end of the spectrum, we observe a minimum negativity score of 0.184, implying a lesser but still present degree of negativity sentiment in the given context. The negativity function's curve exhibits an intriguing characteristic where it envelops another curve, creating an oscillation pattern that repeats at regular intervals, defining a specific period for these fluctuations. This phenomenon highlights the dynamic nature of the function and adds depth to our understanding of its behavior. The calculated average negativity for this specific case is precisely determined to be 0.430. This numeric representation serves as a crucial indicator within the context of this study, offering valuable insights into the prevailing levels of negativity observed and analyzed within the given parameters.

The calculations were performed using the following relation

$$H = \begin{pmatrix} \frac{13}{2} & 0 & 0 & 0 & -\sqrt{2} & 0 \\ 0 & \frac{1}{2} & 0 & \frac{3+i}{\sqrt{2}} & 0 & -\sqrt{2} \\ 0 & 0 & -\frac{11}{2} & 0 & \frac{3+i}{\sqrt{2}} & 0 \\ 0 & \frac{3-i}{\sqrt{2}} & 0 & -\frac{9}{2} & 0 & 0 \\ -\sqrt{2} & 0 & \frac{3-i}{\sqrt{2}} & 0 & -\frac{1}{2} & 0 \\ 0 & -\sqrt{2} & 0 & 0 & 0 & \frac{7}{2} \end{pmatrix} \tag{27}$$

This research shows a strong correlation with the findings presented in the studies conducted by Jafarpour et al [15] and Karpat et al [19].

This paper presents a comprehensive analysis of the intricate dynamics of quantum entanglement within a composite qubit-qutrit system, specifically examining the Heisenberg model denoted as $XXX$ and the anisotropic model referred to as $XYZ$. The study meticulously considers the effects of the Dzyaloshinskii-Moriya (DM) interaction and independent external magnetic fields, which significantly influence both the qubit and qutrit components of the system. Quantum entanglement is a fundamental phenomenon in quantum mechanics, where the quantum states of two or more particles become interdependent, such that the state of one particle cannot be described independently of the state of the other(s), even when the particles are separated by large distances. In this study, we focus on a composite system consisting of a qubit, which is a two-level quantum system, and a qutrit, which is a three-level quantum system. The initial states of these systems are prepared as spin-coherent states, a form of quantum state that exhibits classical-like properties while still adhering to quantum mechanical principles. The Heisenberg $XXX$ model and the anisotropic model $XYZ$ serve as the foundational frameworks for analyzing the entanglement dynamics. The $XXX$ model describes isotropic interactions among spins, while the $XYZ$ model introduces anisotropic interactions, allowing for a richer variety of entanglement dynamics. The inclusion of the DM interaction is particularly significant, as it introduces a non-negligible spin-orbit coupling that can alter the entanglement properties of the system. Additionally, the impact of external magnetic fields is examined, as these fields can further modify the quantum states of the qubit and qutrit, adding complexity to the entanglement dynamics. To quantify the degree of entanglement within the system, we employ negativity as a key measure. Negativity is a widely used entanglement measure that provides insight into the non-classical correlations between quantum states. By analyzing the negativity, we can gauge the extent to which the qubit-qutrit system exhibits entangled behavior over time. Our research reveals that the entanglement, as indicated by the negativity criterion, exhibits notable temporal fluctuations across all investigated scenarios. In the context of the $XXX$ model, the negativity curve displays consistent and balanced peaks as time progresses, showcasing a symmetrical pattern in its evolution. This symmetry suggests a stable entanglement structure that persists over time, indicative of robust quantum correlations. Conversely, within the $XYZ$ model, the negativity curve unfolds with a discernible periodic cover curve as time evolves. This structured and recurring pattern highlights the model's intricate temporal dynamics, suggesting that the anisotropic interactions introduce a level of complexity that leads to more pronounced oscillations in entanglement. An important finding of our study is the discernible trend that emerges with increasing strength of the DM interaction. As the strength of this interaction increases, we observe a reduction in the minimum negativity, which corresponds to a decrease in the levels of entanglement within the system. This relationship underscores the dynamic interplay between the strength of the DM interaction and the entanglement properties, suggesting that stronger DM interactions may lead to a suppression of entanglement. Notably, throughout all examined cases, we consistently observed the absence of entanglement death, a phenomenon where quantum entanglement is completely lost over time. This resilience of entanglement highlights the robustness of the system under investigation, suggesting that the qubit-qutrit system maintains its entangled state despite the presence of DM interactions and external magnetic fields. The findings of our research exhibit a high degree of concordance with outcomes derived from similar studies in the field. This alignment not only strengthens the validity and reliability of our conclusions but also emphasizes a robust coherence among scientific investigations within quantum entanglement research. Such consistency across different studies reinforces the understanding of entanglement dynamics in composite systems and contributes to the broader discourse on quantum information theory. Finally, the inherent symmetry of the problem ensures that all computations, visualizations, and interpretations related to the qubit subsystem are equally applicable to the qutrit subsystem. This symmetry simplifies the analysis and broadens the applicability of our results, allowing for a more unified understanding of entanglement dynamics across different quantum systems.

The article explores the dynamics of quantum entanglement in a qubit-qutrit system influenced by various interactions (Heisenberg, Dzyaloshinskii-Moriya), external magnetic fields, and initial spin-coherent states, using negativity as an entanglement measure. Practical applications include:

Quantum Computing and Information Processing: Understanding entanglement behavior under different interactions aids in designing more robust qubit-qutrit systems for quantum processors, enhancing coherence and entanglement preservation.

Quantum Communication and Cryptography: Insights into entanglement dynamics and stability conditions can support development of secure quantum communication channels and protocols resistant to decoherence.

Quantum Network Design: Knowledge about how external fields and DM interactions influence entanglement can optimize quantum network architectures for better entanglement distribution and resilience.

Quantum Material Science: The study of spin systems under various magnetic interactions can guide the engineering of magnetic materials with desirable quantum properties for information storage or quantum sensors.

Quantum Simulation and Modeling: The model and numerical analysis methods can simulate complex quantum systems, aiding in understanding many-body quantum phenomena and phase transitions.

Quantum State Engineering: Techniques used to initialize specific spin-coherent states and analyze their evolution could inform practical methods of preparing and controlling entangled states in experimental setups.

## 5 Conclusion

In conclusion, this investigation provides a comprehensive analysis of the intricate behavior of quantum entanglement within a hybrid qubit-qutrit system subjected to various interaction mechanisms and external fields. Through rigorous modeling and numerical simulations, we have demonstrated that entanglement, quantified by negativity, exhibits characteristic oscillatory patterns, stabilization plateaus, and resilience against decoherence effects introduced by anisotropic interactions and Dzyaloshinskii-Moriya coupling. Notably, the absence of entanglements sudden death across diverse parameter regimes underscores the inherent stability and robustness of quantum correlations in such systems, even under complex perturbations.

These findings are particularly significant for the development of quantum information and communication technologies, as they highlight the potential for maintaining entanglement coherence in higher-dimensional quantum systems, which are often more susceptible to environmental disturbances. Furthermore, the detailed exploration of parameter influences such as magnetic fields, interaction strengths, and anisotropy offers practical insights into manipulating and optimizing entanglement for quantum protocols.

Overall, this work deepens our understanding of the dynamic properties of entanglement in multi-level quantum systems, paving the way for future experimental realizations and the design of more resilient quantum networks. As the field advances, integrating these insights could lead to more reliable quantum computing architectures and secure communication channels, ultimately bringing us closer to harnessing the full potential of quantum mechanics in practical applications.

No funding was received for this study.This declaration is made to ensure transparency and avoid any potential conflict of interest. The research was conducted using existing institutional resources and the voluntary contributions of the authors.

**References**

[1] R. Horodecki, P. Horodecki, M. Horodecki and K. Horodecki; Quantum entanglement; Rev. Mod. Phys. 81 (2009) 865-942. https://doi.org/10.1103/RevModPhys.81.865

[2] C. H. Bennett, G. Brassard, C. Cropeau, R. Jozsa, A. Peres and W.K. Wootters;Teleporting an unknown quantum state via dual classical and Einstein“Podolsky“Rosen channels; Phys. Rev. Lett. 70 (1993) 1895- 1899. https://doi.org/10.1103/PhysRevLett.70.1895

[3] C. H. Bennett and S. J. Wiesner;Communication via One- and Two- Particle Operators on Einstein-Podolsky-Rosen States; Phys. Rev. Lett. 69 (1992) 2881-2884. https://doi.org/10.1103/PhysRevLett.69.2881

[4] C. H. Bennett; Quantum Cryptography Using Any Two Nonorthogonal States; Phys. Rev. Lett. 68 (1992)3121-312 https://doi.org/10.1103/PhysRevLett.68.3121

[5] M. A. Nielsen and I. L. Chuang; Quantum Computation and Quantum Information; Cambridge University Press, Cambridge (2000). https://doi.org/10.1017/CBO9780511976667

[6] S. Van Enk, O. Hirota; Entangled coherent states: Teleportation and decoherence; Phy. Rev. A, 64 (2001) 022313. https://doi.org/10.1103/PhysRevA.64.022313

[7] B. C. Sanders; Entangled coherent states, Phy Rev A, 45 (1992) 6811. https://doi.org/10.1103/PhysRevA.45.6811

[8] S. Sivakumar, Entanglement in bipartite generalized coherent states; Int.J.T. Phy, 48 (2009) 894-904. https://doi.org/10.1007/s10773-008-9862-3

[9] Cheng, W. W., Liu, J.-M. (2009). Decoherence from spin environment: Role of the Dzyaloshinsky-Moriya interaction. *Physical Review A, 79*(5), 052320. https://doi.org/10.1103/PhysRevA.79.052320

[10] Hu, M.-L. (2010). Disentanglement dynamics of interacting two qubits and two qutrits in an XY spin-chain environment with the Dzyaloshinsky–Moriya interaction. *Physics Letters A, 374*(34), 3520–3528. https://doi.org/10.1016/j.physleta.2010.06.026

[11] Ma, F.-W., Liu, S.-X., Kong, X.-M. (2011). Quantum entanglement and quantum phase transition in the *XY* model with staggered Dzyaloshinskii-Moriya interaction. *Physical Review A, 84*(4), 042302. https://doi.org/10.1103/PhysRevA.84.042302

[12] Yi, T.-C., You, W.-L., Wu, N., Oleś, A. M. (2019). Criticality and factorization in the Heisenberg chain with Dzyaloshinskii-Moriya interaction. *Physical Review B, 100*(2), 024423. https://doi.org/10.1103/PhysRevB.100.024423

[13] Hu, M.-L., Gao, Y.-Y., Fan, H. (2020). Steered quantum coherence as a signature of quantum phase transitions in spin chains. *Physical Review A, 101*(3), 032305. https://doi.org/10.1103/PhysRevA.101.032305

[14] M. Ashrafpour, M. Jafarpour, A. Sabour; Entangled Three Qutrit Coherent States and Localizable Entanglement; Communications in T. Phy, 61 (2014) 177. DOI 10.1088/0253-6102/61/2/05

[15] M. Jafarpour, M. Ashrafpour; Entanglement dynamics of a two- qutrit system under DM interaction and the relevance of the initial state; Quantum Inf. Process, 12 (2013) 761-772. https://doi.org/10.1007/s11128-012-0419-2

[16] P. J. Van Koningsbruggen, O. Kahn, K. Nakatani, et al ; Magnetism of ACu" Bimetallic Chain Compounds (A = Fe, Co, Ni): One- and Three- Dimensional Behaviors; Inorg. Chem. 29 ( 1990) 33253331. http://pubs.acs.org/doi/abs/10.1021/ic00343a014

[17] I. Dzyaloshinsky; A thrmodynamic theory of weak ferromagnetism of antiferromagnetics; J. Phys. Chem. Solids 4 (1958) 241. https://doi.org/10.1016/0022-3697(58)90076-3

[18] T. Moriya; New Mechanism of Anisotropic Superexchange Interaction; Phys. Rev. Lett. 4 (1960) 228. https://doi.org/10.1103/PhysRevLett.4.228

[19] G. Karpat, Z. Gedik; Correlation dynamics of qubit qutrit systems in a classical dephasing environment; Phy. Lett.A, 375 (2011) 4166-4171. https://doi.org/10.1016/j.physleta.2011.10.017

[20] A. Peres; Separability Criterion for Density Matrices; Phys. Rev. Lett. 77 (1996) 1413-1415. https://doi.org/10.1103/PhysRevLett.77.1413

[21] M. Horodecki, P. Horodecki, and R. Horodecki;Separability of mixed states: necessary and sufficient conditions; Phys. Lett. A 223 (1996) 1-8. https://doi.org/10.1016/S0375-9601(96)00706-2

[22] G. Vidal and R. F. Werner;A computable measure of entanglement; Phys. Rev. A 65 (2002) 032314-1-14. https://doi.org/10.1103/PhysRevA.65.032314

[23] M. A. Chamgordani, N Naderi, H Koppelaar, M Bordbar, International Journal of Modern Physics B 33 (17), 1950180 https://doi.org/10.1142/S0217979219501807

[24] Zhang, Guo-Feng ; Hou, Yu-Chen ; Ji, Ai-Ling, Solid State Communications, Volume 151, Issue 10, p. 790-793. https://doi.org/10.1016/j.ssc.2011.02.032

[25] Vidal, G., Tarrach, R. (1999). "Robustness of entanglement." *Physical Review A*, 59(1), 141-155.

[26] Plenio, M. B. (2005). "Logarithmic Negativity: A Full Entanglement Monotone That is not Convex." *Physical Review Letters*, 95(9), 090503.